\documentstyle [12pt,a4p,epsfig,amsmath,multicol]{article}
\begin{document}
\vspace*{0.6cm}

\begin{center} 
{\normalsize\bf Invariant space-time intervals, clock synchronisation and correct application 
  of Lorentz transformations in special relativity}
\end{center}
\vspace*{0.6cm}
\centerline{\footnotesize J.H.Field}
\baselineskip=13pt
\centerline{\footnotesize\it D\'{e}partement de Physique Nucl\'{e}aire et 
 Corpusculaire, Universit\'{e} de Gen\`{e}ve}
\baselineskip=12pt
\centerline{\footnotesize\it 24, quai Ernest-Ansermet CH-1211Gen\`{e}ve 4. }
\centerline{\footnotesize E-mail: john.field@cern.ch}
\baselineskip=13pt
\vspace*{0.9cm}
\abstract{ Space-like and time-like invariant space-time intervals are used to
 analyse measurements of spatial and temporal distances. The former are found to be
 Lorentz invariant --there is no `relativistic length contraction', whereas the
 latter exhibit the experimentally confirmed `time dilatation' effect, without, however,
 `relativity of simultaneity'. It is shown how the spurious predictions of length
  contraction and relativity of simultaneity arise from the use of inconsistent 
   synchronisation procedures for spatially separated clocks.}
\vspace*{0.9cm}
\normalsize\baselineskip=15pt
\setcounter{footnote}{0}
\renewcommand{\thefootnote}{\alph{footnote}}

 With a particular choice of coordinate axes and clock synchronisation convention, to be discussed
    below, the space-time Lorentz Transformation\footnote{The Lorentz Transformation (1)-(5)
   was first written down in the Literature by Larmor in 1900~\cite{Larmor}. See ~\cite{Kittel}
   for a discussion of the historical priority issues.  It was
   independently derived by Einstein~\cite{Ein1} and given by Poincar\'{e}~\cite{Poincare} in 1905.
   The appellation `Lorentz Transformation' is due to Poincar\'{e}. In Lorentz's last
   pre-relativity paper~\cite{Lorentz} another space-time transformation was given, from which
   the Lorentz Transformation (1)-(5) may be obtained by the substitution: $x \rightarrow x-vt$.}
 (LT) may be written as:
\begin{eqnarray}
    x' & = & \gamma[x-vt] \\
    t' & = & \gamma[t-\frac{v x}{c^2}] \\
    y'  & = & y \\
   z'  & = & z
\end{eqnarray}
 where
 \begin{equation}
 \gamma\equiv \frac{1}{\sqrt{1-(\frac{v}{c})^2}}  
\end{equation}
 In (1)-(4) the space-time coordinates $x,y,z,t$ (E) and  $x',y',z',t'$ (E')
 specify the same event as observed in two inertial frames S and S' respectively. The frame S' is in
  uniform motion relative to
 S with velocity $v$ in the direction of the positive $x$-axis. The $x$,$x'$-axes and the $y$,$y'$-axes 
 are parallel and $c$ is the speed of light in vacuum. 
 \par It was shown by Minkowski~\cite{Mink} that with every pair of events, $E$ and $E_0$ in S, and 
 $E'$ and $E_0'$ in S', with space-time coordinates connected by the  LT (1)-(5), may be associated space-like 
  and time-like invariant intervals $\Delta s$, $\Delta \tau$, respectively, 
  according to the definitions:
\begin{eqnarray}
 (\Delta s)^2 & \equiv & -c^2(\Delta \tau)^2  \nonumber \\
  & \equiv & (\Delta x)^2 +(\Delta y)^2 +(\Delta z)^2 - c^2(\Delta t)^2 
 \nonumber \\
   & = & (\Delta x')^2 +(\Delta y')^2 +(\Delta z')^2 - c^2(\Delta t')^2
 \end{eqnarray}
  where $\Delta x \equiv x-x_0$, $\Delta t \equiv t-t_0$ etc. Suppose now that $E'$ and $E'_0$
  are events on the world line of a clock, A', that is at rest in S' at some fixed position 
  on the $x'$ axis. Since, in this case, $\Delta x' = \Delta y'= \Delta z'= 0$,
  (6) may be written as:
  \begin{equation}
  c^2(\Delta \tau(A'))^2 = c^2(\Delta t(A'))^2 -(\Delta x(A'))^2
   = c^2(\Delta t'(A'))^2
   \end{equation}  
    Since the clock A' moves with uniform velocity $v$ in the positive $x$-direction in S:
   \begin{equation}
     \Delta x(A') = v \Delta t(A')
    \end{equation}
    Substituting (8) in (7) gives:
   \begin{equation}
    c^2(\Delta \tau(A'))^2 = c^2(\Delta t(A'))^2(1-\frac{v^2}{c^2})
     = \frac{c^2(\Delta t(A'))^2}{\gamma^2} =  c^2(\Delta t'(A'))^2
   \end{equation}
   or
     \begin{equation}
   \Delta t(A') = \gamma\Delta t'(A') \equiv \gamma\Delta \tau(A')
     \end{equation}
     This equation describes the time dilatation effect of SR ---the clock A' appears
     to be running slower, to an observer in the frame S, by a factor $1/\gamma$, as compared
     to the observed rate of an identical clock at rest in the frame S. The notation $\tau(A´) =  t'(A')$
      in (10) shows that, in this equation,  $t'(A')$ is a proper time interval of the clock A' whereas
      $t(A')$ is the corresponding time interval as recorded by a clock at rest in the frame S, relative
      to which, A' is in motion. 
 \par Consider now a second clock, B', also at rest on the $x'$ axis in S'. In terms of clock
   times the time dilatation formulae for the clocks A' and B' are:
  \begin{eqnarray}
  t(A')- t_0(A') & = & \gamma[\tau(A')- \tau_0(A')] \\
  t(B')- t_0(B') & = & \gamma[\tau(B')- \tau_0(B')]
  \end{eqnarray}
   The parameters  $\tau_0(A')$ and $\tau_0(B')$ specify the synchronisation of the clocks
    A' and B' that record the times  $\tau(A')$ and $\tau(B')$, whilst  $t(A')$,$t(B')$,$t_0(B')$
    and $t_0(B')$ are the recorded times and synchronisation constants of a clock or clocks at rest in S.
    All the times in (11) and (12) are those seen by observers at rest in S. Therefore, in order
   to actually measure the time dilatation effect of Eqn(10), at least one clock at rest in S, as well as a moving
   clock, such as A' or B', is necessary. Suppose now that the clocks A' and B' are stopped and set
    to the common value:
    \begin{equation}
 \tau_0(A') = \tau_0(B') \equiv \tau_0(A',B') 
   \end{equation}
    Two marks, separated by the distance, $L$, are made along the $x$-axis so that the moving clocks
    A' and B' arrive simultaneously at the marks as seen by an observer at rest in S. At this instant
   the clocks A' and B' are started. It is easily shown by use of the LT (1)-(5) with the coordinate
   origins situated at the positions of the marks (say $x$, $x'$ for the mark coincident
    with A', and $x \rightarrow X$, $x' \rightarrow  X'$ for the mark coincident with B',
    where $x = X + L$ and $x' = X' + L'$)
    that the clocks are observed to be synchronous in both S and S'.
     The use of the LT (1)-(5) with the coordinate systems ($x$,$x'$)
    for A' and ($X$,$X'$) for  B' corresponds to the choice  $t_0(A') = \tau_0(A') = t_0(B') = \tau_0(B') = 0$
      for the initial clock settings. 
     The parameters $t_0(A')$ and $t_0(B')$ are actually the initial 
   settings of clocks in S, which may conveniently be aligned with the marks along the $x$-axis.
  Without loss of generality,
  they may be chosen to be the same as the clock settings of A' and B', respectively,
   at the instant they are started. It then follows that:
 \begin{equation}
   t_0(A') = \tau_0(A') = \tau_0(A',B') = t_0(B') =  \tau_0(B') \equiv  T_0(A',B') 
  \end{equation}
   Where $T_0(A',B')$ is the time recorded by local synchronised clocks in both S and S' at the instant
   when A' and B' are started.
    The only difference between the clock synchronisation procedure just described and that which 
   is conventionally used in special relativity, is that it is performed, at the same instant 
    in S, simultaneously for two, spatially separated, clocks A' and B', instead of for a single
    clock. Note that (14) denotes a {\it particular choice} of the {\it arbitary} parameters
    $t_0(A')$, $\tau_0(A')$, $t_0(B')$ and $\tau_0(B')$. In particular, these parameters are {\it not
    connected} by LT equations.
  \par After synchronisation, (11) and (12) are then written as:
  \begin{eqnarray}
  t(A')- T_0(A',B') & = & \gamma[\tau(A')- T_0(A',B')] \\
  t(B')- T_0(A',B') & = & \gamma[\tau(B')- T_0(A',B')]
  \end{eqnarray}
 It follows from these equations that all events which are simultaneous in S' at the positions
  of the clocks A' and B': $\tau(A') = \tau(B') \equiv \tau(A',B')$ are also observed to be
   simultaneous in S:
 \begin{equation}
  t(A') = \gamma[\tau(A',B')- T_0(A',B')]+T_0(A',B') =  t(B')
   \end{equation}
   A quite general  consequence of this equation is that events which are judged to be simultanous
   in S' according to local synchronised clocks are also observed to be simultaneous by such clocks in S
    ---there is no `relativity of simultaneity'.
  \par Suppose now that the separation of the clocks A' and B' is $L'$, i.e. $L' \equiv x'(B')-x'(A')$.
 A space-like Lorentz-invariant interval $\Delta s(B',A')$ may be defined, using the LT (1)-(5) as:
  \begin{equation}
   (\Delta s(B',A'))^2 \equiv  (\Delta x(B',A'))^2 -c^2 (\Delta t(B',A'))^2
    = (L')^2-c^2 (\Delta \tau(B',A'))^2
   \end{equation}
   where
  \begin{equation}
  \Delta t(B',A') \equiv t(B')-t(A'),~~~~~\Delta \tau(B',A') \equiv \tau(B')-\tau(A')
  \end{equation}
   In the frame S, the clocks are moving with a fixed separation, $L$, in the direction
   of the positive $x$-axis. This distance is measured by noting the positions of the clocks
   at some instant, $t$, in the frame S and subtracting the corresponding $x$-coordinates:
   \begin{equation}
    L \equiv x(B',t)- x(A',t) \equiv \Delta x(B',A')
  \end{equation}
  where $t = t(B') = t(A')$ or
 \begin{equation}
 \Delta t(B',A') = 0
   \end{equation}
   When $t(B') = t(A')$, (15) and (16) give  $\tau(B') = \tau(A')$ or
  \begin{equation}
 \Delta \tau(B',A') = 0
   \end{equation}
  Substituting (20),(21) and (22) in (18), and taking the positive square root gives:
  \begin{equation}
  \Delta s(B',A') = L = L'
  \end{equation}  
   Contrary, then, to what is claimed in the usual interpretation of special relativity, the distance between 
   two objects that are at rest in some inertial frame is a Lorentz invariant quantity --there is no
   `length contraction'. 

    By considering the time dilatation effect, as observed in the frame S' when S moves with speed $v$ along
    the negative $x´$-axis, the following formulae, analogous to (15) and (16), are obtained.
  \begin{eqnarray}
  t'(A)- T_0(A,B) & = & \gamma[\tau(A)- T_0(A,B)] \\
  t'(B)- T_0(A,B) & = & \gamma[\tau(B)- T_0(A,B)]
  \end{eqnarray}
     \par  The parameters $T_0(A',B')$ in(15),(16) and $T_0(A,B)$ in (24) and (25) are the readings of each pair of synchronised clocks
   (A' and B' in S, A and B in S) at the instant of synchronisation and may be freely chosen
   without changing any physical predictions. As a last step these two parameters can also be chosen to be equal,
    defining in this way {\it the same synchronisation convention} for all four clocks.
  This means that:
   \begin{equation}
     t_0(A',B') = \tau_0(A',B') \equiv T_0(A',B') \equiv  T_0
   \end{equation}
     and 
:  \begin{equation}
    t_0(A,B) =  \tau_0(A,B) \equiv T_0(A,B) \equiv T_0 
   \end{equation}   
    (26) and (27) imply that (15),(16), (25) and (26) become, after synchronisation of A with A'
    and B with B':
  \begin{eqnarray}
     t(A') -  T_0 & = & \gamma[\tau(A')- T_0] \\
     t(B') -  T_0 & = & \gamma[\tau(B')- T_0] \\
     t'(A) -  T_0 & = & \gamma[\tau(A)-  T_0]  \\
     t'(B) -  T_0 & = & \gamma[\tau(B)-  T_0]
  \end{eqnarray}
    \par The results just obtained concerning distance
    and time interval measurements were derived solely from the operational definitions of such measurements
   and the Lorentz-invariant interval  relation (6). No particular coordinate system was specified.
   The results will now be rederived using the LT. This calculation will reveal why the usual
   `length contraction' and `relativity of simultaneity' predictions of conventional SR are
     erroneous.
    \par The standard space-time LT of (1)-(5) corresponds to a particular choice of coordinate
    axes and clock synchronisation convention. For an arbitary choice of these the LT is written as:
\begin{eqnarray}
    x'-x'_0 & = & \gamma[x-x_0-v(t-t_0)] \\
    t' -t'_0 & = & \gamma[t-t_0-\frac{v(x-x_0)}{c^2}] \\
    y'-y'_0  & = & y-y_0 \\
   z'-z'_0  & = & z-z_0
 \end{eqnarray} 
     The `standard' LT of (1)-(5) corresponds to the choice $t_0 = t'_0 = 0$ for the synchronisation
    convention and $x_0 = x'_0 = y_0 = y'_0= z_0 = z'_0$ for the choice of coordinate axes. 
     Thus when $x'= t' = 0$ also  $x = t = 0$. Placing the clock A' at the origin of the
     x' axis and using the standard LT give, for events on the world line of A', the transformations:
 \begin{eqnarray}
   x'(A') & = &\gamma[x(A')- v t(A')] = 0 \\
   t'(A') \equiv \tau(A') & = & \gamma[t(A')- \frac{v x(A')}{c^2}] \\
    y'(A') & = &  y(A') =  z'(A')=   z(A') = 0 
 \end{eqnarray}
   Using (36) to eliminate $x(A')$ in (37) gives, after use of (5) and transposition
\begin{equation}  
   t(A') =  \gamma  \tau(A')
\end{equation}   
   which is just the time dilatation relation (28) for the clock A', with the choice $\tau_0 = 0$ for the clock
    setting at the time of synchronisation.
    Using the same synchronisation convention as for A' the LT of events on the world
    line of the clock B' {\it when it is synchronised with} A' { \it in} S' is:
\begin{eqnarray}
   x'(B')-L & = &\gamma[x(B')-L- v t(B')] = 0 \\
 t'(B')  \equiv \tau(B') & = & \gamma[t(B')- \frac{v (x(B')-L)}{c^2}] \\
    y'(B') & = &  y(B') =  z'(B')=   z(B') = 0 
 \end{eqnarray}
 Setting $t(B´)= 0$, $x(B') = L$ in (40) and (41) gives $t'(B') = 0$. In consequence
  $t'(A´)= t'(B´)= t(A´)= t(B´) = 0$ when  $x(A') = 0$ and $x(B') = L$, so that A' and B' are correctly
   synchronised by (36),(37),(40) and (41).
    Using (40) to eliminate $ x(B')$ in (41), gives similarly to (39):
 \begin{equation}  
   t(B') =  \gamma  \tau(B')
\end{equation}
  in agreement with (29) for $\tau_0 = 0$.   
  The LT (36),(37) may be more simply written as:
 \begin{equation}
   x'(A') =  0,~~~~x(A') = v t,~~~~  t = \gamma \tau'
 \end{equation}   
   and that in (40),(41) as
 \begin{equation}
   x'(B') =  L,~~~~x(B') = v t+L,~~~~  t = \gamma \tau'
 \end{equation}  
  where 
 \begin{equation}
   t(A') = t(B') \equiv t,~~~~  \tau(A') =  \tau(B') \equiv \tau' 
 \end{equation}   
   Here $\tau'$ is the proper time of all synchronised clocks at rest in  S', while
   $t$ is the corresponding (time-dilatated) time recorded by a similar clock at rest in S.
   Inspection of (44) and (45) shows that $x'(B') - x'(A') \equiv L' = x(B') - x(A') \equiv L$,
    for all values of t, in accordance with the Lorentz invariant relation (23).
    \par The standard text-book derivation of the `length contraction' and `relativity
    of simultaneity' effects is now presented and discussed. What is done is to apply
    the standard LT (1)-(5) not only to the clock A' but also to B'. When this is
    done, in place of (40) and (41), the following equations are obtained:
 \begin{eqnarray}
   x'(B') & = &\gamma[x(B')- v t(B')] \\
   \tau(B') & = & \gamma[t(B')- \frac{v x(B')}{c^2}] 
 \end{eqnarray}
   Subsituting $t(A') = 0$ in (36),(37) and  $t(B') = 0$ in  (47),(48)  gives
   \begin{eqnarray}
    x'(A') & = & x(A') =  0 \\
    \tau(A') & = & 0 \\
   x'(B') & = &\gamma x(B')   \\
   \tau(B') & = & - \gamma \frac{v x(B')}{c^2} \ne 0 =  \tau(A') 
 \end{eqnarray}
    Equation (51) may be re-witten as:
 \begin{equation}
   x'(B')- x'(A')  =   \gamma[ x(B')-x(A')]
 \end{equation}
  or
 \begin{equation}
   L'  =  \gamma L
 \end{equation}
   where $L \equiv x(B')-x(A') $ is the distance between A' and B' in S at $t = t(A') =  t(B') = 0$,
   according to (36),(37),(47) and (48). This is the `length contraction' effect of conventional
   special relativity.
   Equation (52) is interpreted as `relativity of simultaneity' --- since $ \tau(B') \ne \tau(A')$,  events which are
   simultaneous in S are not so in S'. What it really shows however is that { \it the
   use of the LT (1)-(5) for the clock} B' {\it at} $x' = L$  {\it does not synchronise the clock
    pair} B-B' { \it in the same way as the pair} A-A'. In fact, this ansatz uses  different
    synchronisation conventions for the clock pairs A-A' and B-B'. To understand this,
    introduce the clocks A and B in S, coincident along the x-axis with A' and B' respectively,
    at the instant when $t(A') = t(A) = t(B') = t(B) = 0$. For the clock A', (39) gives
     $\tau(A') = t(A')/\gamma = 0$. The LT (1)-(5) applied to A' then corresponds to the
     synchronisation convention of (28) for the clock pair A-A'  with $T_0 = 0$. i.e.
 \begin{equation}
   t_0(A)= \tau_0(A') = T_0 = 0 
 \end{equation}  
      If the clocks A and B are synchronised in S then
    $t_0(B) = t_0(A) = 0$. However, according to (51), (52) and (12)
     $\tau_0(B') = -v L'/ c^2$. The synchronisation convention for the clock 
    pair B-B' is then:
 \begin{equation}
   t_0(B) =  \tau_0(B')+ \frac{v L'}{c^2} = T_0 = 0 
 \end{equation} 
   which is evidently different to that of the clock pair A-A' in (55).
   Actually, comparison of the (55) and (56) shows that with the synchronisation convention
  of (56) {\it unlike the clocks} A {\it and} B {\it in} S, {\it the clocks} A' {\it and} B' {\it are not
   synchronised in the frame} S '. This (and only this) is the origin of the `relativity of simultaneity'
   effect of conventional
   special relativity where the standard form (1)-(5) of the LT is universally employed for clocks at arbitary
   spatial locations in the frame S' 
  \par Use of (43), following from the LT (40)-(41) and (12), shows instead that $\tau_0(B') = 0$.
   The synchronisation of the clocks A and B in S gives 
   $t_0(B) = t_0(A) = 0$ 
  so that instead of (56) the synchronisation
   convention for the clock pair B-B' becomes:
 \begin{equation}
  t_0(B) = \tau_0(B') = T_0 = 0 
 \end{equation}  
  identical with that for the clock pair A-A' in (55). In this case, unlike in (56), the clocks A' and B' are
   synchronised in the frame S' and the clocks A and B  are synchronised in the frame S.    
  \par The essential message of the present paper is then that, in an operational
  sense, i.e. in its application to experimental physics, the LT does not primarily 
  describe `space-time geometry' but rather how the times displayed by synchronised clocks
  appear to observers in different inertial reference frames. Inspection of (44) and (45)
   shows that there is a universal time dilatation effect for clocks at rest in S' and in motion
    in S but that the equations containing spatial coordinates are the same as for
   a Galilean space-time transformation, given by the $c \rightarrow \infty$ limit of (1) and (2).
    Special relativity therefore changes the classical concept of time but not that of space. A more
   detailed study
  of this subject, including rigorous experimental definitions for measurements
  of spatial and temporal intervals, as well as suggestions for several
  practical clock synchronisation procedures, not making use of light signals,
  is presented in Ref.~\cite{JHFSTP1}. 
  \par An immediate application of the above results is to the analysis of the train/embankment
   gedankenexperiment~\cite{EinTEG} proposed by Einstein to introduce the concept
   of `relativity of simultaneity' before any mention of the LT. Two lightning strokes
   hit the embankment at two points that coincide with the front and the end of the moving train
   at that instant. There are observers mid-way between the points on the embankment
   and at the middle of the train. The question is asked whether these observers will
   judge the strokes to be simultaneous. It is clear from the time transformation
   equations in the LT (44) and (45) that if the strokes are simultaneous in the embankment
   frame they will be observed to be also simultaneous in the train frame, so that both observers
    must judge the strokes to be simultaneous on receiving light signals generated by the strokes.
    Why the analysis of these light signals lead Einstein
    to the erroneous conclusion that the strokes would not be judged simultaneous 
    by the observer on the train is discussed elsewhere~\cite{JHFTEG}.
    \par In order to use the same synchronisation convention for both pairs of clocks 
      with one in S' and the other in S, the general prescription is always to use
    a `local' LT, i.e. one in which the coordinate origin of the frame in which the
    clock is at rest is at the position of the clock. Such a local LT is defined
    for the clock A' discussed above by the LT (36)-(38). The LT in (40) and (41)
    is converted to a local one by the coordinate transformations: $x-L \rightarrow X$
    and $x'-L \rightarrow X'$. For the clock B'  for which $x'(B') = L$, $X'(B') = 0$ so that
    the coordinates $(X,t)$, $(X',t')$ are local ones for the clock B'.
    In fact, after these coordinate transformations,
    (40) and (41) are identical to the 'standard' transformations (1) and (2) with the replacements
    $x \rightarrow X$ and $x' \rightarrow X'$. An extensive discussion of the physical
    motivation and use of the local LT is presented
    in Ref.~\cite{JHFLLT}. Unlike in some applications of the standard LT, translational
    invariance is always respected and some causal paradoxes~\cite{Soni,JHFCP} of conventional special
    relativity are resolved.
    \par Satellite borne experiments to test, for the first time, the existence (or not)
    of the `relativity of simultaneity' effect arising from the exclusive use of the
    standard LT (1)-(5) are proposed in Ref.~\cite{JHFRSE}.
  \par{\bf Added Note}
     \par The argument presented in (18)-(23) for the absence of length contraction is erroneous, although the
      last member of (23), $L = L'$ is correct. It is wrongly assumed, in deriving (18), that the clocks A'
      and B' are synchronised  when they are both described by the LT (1)-(5). If A' is situated at the
      origin of coordinates in S' and B' at $x'= L$, the correct transformation equations are given by
       (36)-(38) for A' and (40)-(42) for B'.  In consequence it is found, instead of (18), that (dropping
       the labels A' and B')
        \begin{eqnarray}
             (\Delta x')^2-c^2 (\Delta \tau)^2 & = &(\Delta x)^2 -c^2 (\Delta t)^2
    \nonumber \\
              & &+2L^2 -2 L\Delta x +2\gamma (\Delta x-L) -2 \gamma v \Delta t
   \end{eqnarray}
     In view of (15) and (16), setting $\Delta t = 0$ implies $\Delta \tau = 0$, Since 
       $\Delta x~=~L$ when $\Delta t = 0$, (58) gives
       \begin{equation}
 \Delta x' \equiv L' = L
      \end{equation}
       Contrary, however, to the statement in the last paragraph on page 4, the relation $L = L'$ has
      {\it not} been derived from a space-like invariant interval relation without consideration
       of a specific spatial coordinate system. In fact strict attention to the meaning of the
       spatial coordinates describing a synchronised clock, as in (40) and (41), is essential to
      show that $L = L'$ and that the LC effect of conventional special relativity is spurious.
         \par In the relation, given by replacing $L$ on the left side of (40) by $L'$:
   \begin{equation} 
       x'(B')-L'  =  \gamma[x(B')-L- v t(B')] = 0
 \end{equation}
 both $L$ and $L'$ are constants which are independent of $v$ since
   \begin{equation} 
    L = \left. x(B')\right|_{t(B') = 0}
 \end{equation} 
   The transformation equation (60) is therefore valid for all values of $v$ with the same
   choice of spatial coordinate systems in S and S'. In particular, it is valid when $v = 0$, $\gamma = 1$ 
    and $x \rightarrow x'$: 
  \begin{equation} 
     x'(B')-L'  =  x'(B')-L 
 \end{equation}
   from which it follows that  $L' = L$.
 \par{\bf Acknowledgements}
   \par I would like to acknowledge discussions with B.Echenard, and correspondence with M.Gr\"{u}newald,
   Y.Keilman and D.Utterback, concerning an earlier version of this paper. Although they disagreed with its
   conclusions, their pertinent criticisms have enabled me to improve the clarity of the presentation
   in this version.

\end{document}